\documentclass[aps,prb,showpacs,twocolumn]{revtex4-1}

\usepackage{amssymb}
\usepackage{epsfig}
\usepackage{graphicx}
\usepackage{amsmath}
\usepackage{array,color}
\usepackage{dcolumn}
\usepackage{bm}

\begin{document}

\title{Cellular dynamical mean-field theory study of an interacting topological honeycomb lattice model at finite temperature}

\author{Yao-Hua Chen$^{1}$}

\author{Hsiang-Hsuan Hung$^{2}$}

\author{Guoxiong Su$^{1}$}

\author{Gregory A. Fiete$^{2}$}

\author{C. S. Ting$^{1}$}

\affiliation{
$^1$Texas Center for Superconductivity and Department of Physics,
University of Houston, Houston, Texas 77204, USA
}

\affiliation{
$^2$Department of Physics,
The University of Texas at Austin, Austin, Texas 78712, USA
}

\date{\today}

\begin{abstract}
Topological phases originating from spin-orbit coupling have attracted great attention recently.  In this work, we use cellular dynamical mean field theory with the continuous-time quantum Monte Carlo solver to study the Kane-Mele-Hubbard model supplemented with an additional third-neighbor hopping term.  For weak interactions, the third-neighbor hopping term drives a topological phase transition between a topological insulator and a trivial insulator, consistent with previous fermion sign-free quantum Monte Carlo results [H.-Hung {\it et al.} Phys. Rev. B {\bf 89}, 235104 (2014)].  At finite temperatures, the Dirac cones of the zero temperature topological phase boundary give rise to a metallic regime of finite width in the third-neighbor hopping.  Furthermore, we extend the range of interactions into the strong coupling regime and find an easy-plane anti-ferromagnetic insulating state across a wide range of third-neighbor hopping.  In contrast to the weak coupling regime, no topological phase transition occurs at strong coupling, and  the ground state is a trivial anti-ferromagnetic insulating state.   A comprehensive finite temperature phase diagram in the interaction-third-neighbor hopping plane is provided.
\end{abstract}


\maketitle

\section{Introduction}
Topological insulating states, such as the topological band insulator (TBI), the topological Mott insulator, and other interacting varieties of topological states have been attracted much interest in condensed matter physics\cite{prl0950146802, prl0950226801,prl097036808, rmp08301057, rmp08203045, science034006129,natphys06376, prb083165112,prl0105246809,prb083205101,prl0109066401,prl0108046401,prb088241101,prl0112016404}. These topological insulating states are characterized by topological numbers, such as the Chern number, mirror Chern number, and the $Z_2$ number\cite{science032605967,natcomm03982, prl0110156403, prl01000156401, prl0108026802, nph08067,  prl0990236809, naturephysics050438, prl01090186805, prl0107076801, prl01020256403}. The TBI have been experimentally found in many materials, such as $Bi_2Se_3$, and $HgTe/CdTe$ quantum wells\cite{science031401757, science03250178, prl01060156402, nature04520970, naturematerials090541, naturephysics050398, prl01050076802, prl01060257004}. Besides the experimental progress in the detection of topological insulating states, much theoretical research has been devoted to  the role of lattice geometry on the topological insulating states, including the honeycomb, square, kagome, and more unusual lattices \cite{prb0870155112, prl0109205303, prb0850205102, prb082085106, prb082075125, prb084155116,physe044845}.  In addition to geometric factors,  topological phase transitions can also be induced by a staggered on-site energy\cite{prl0950146802}, Rashba spin-orbit coupling\cite{science031401757, prl0950146802, prl0950226801, prl097036808, prb0890165135}, and a third-neighbor hopping in non-interacting models \cite{ prb0870121113r,prb0890235104}.

Recently, the influence of electronic correlations on topological states has been the focus of many studies.
In the strong coupling limit, interactions could induce magnetic ordering which either breaks the time reversal symmetry, which then spoils the TBI state\cite{prl0106100403,prb085115132,prb0840205121,prb0850205102,prl0107010401,prb0820075106,prb0830205122,prb0850045123,prb0870085109}, or coexists with the topological phases to form an anti-ferromagnetic topological insulator\cite{prb0810245209,prb0880085406,prb0870085134,prb0870195133}. It is also interesting to investigate the topological phase transitions \cite{prb0870121113r,prb0890235104,prb0870205101,mpl0280143001, prb0880195130} at strong interactions and how finite temperatures influence topological states \cite{prb0850125113,jpcm0260175601}.
In particular, interactions and thermal fluctuations have been proposed to drive a nontrivial TBI or otherwise change topological properties\cite{prb082075125,prb0870085109,prl01000156401,epl098057001,jpcm0260175601}.
Thus, it is highly desirable to investigate the effect of interactions on topological systems, particularly at finite temperature, which is important and relevant to real materials.

Many analytical and numerical methods have been developed to investigate interacting systems in the past few years \cite{prl01040116402, prl01010106401, prl01090196402}, among them dynamical mean-field theory (DMFT) is an especially powerful method capable of capturing the Mott transition \cite{rmp068013}.  While single-site DMFT has been shown to work well in three dimensional systems (it is exact in the limit of infinite spatial dimensions), in two-dimensional systems non-local correlations and spatial fluctuations can have an important influence on the physics. To improve the predictions of DMFT, particularly in two-dimensions, cellular dynamical mean-field theory (CDMFT) \cite{rmp07701027, rmp0780865, prl01010186403, prl0100256403, prl0920226402, prl01000056403, prl01050065301, prb0850205102,prb0670075110,prl01080246402} has been developed to incorporate spatially extended correlations. In CDMFT, the original lattice is mapped to an effective cluster impurity model coupled to an effective medium.  An important impurity solver in CDMFT uses the continuous time quantum Monte Carlo method (CTQMC) \cite{prb072035112, rmp0830349}, which is more accurate than the ``traditional" discrete-time QMC method. The momentum-dependent spectral function can be used to detect the different characters of the edge states appearing in the topologically trivial and non-trivial states. In addition, topological phase transitions can also be studied by observing the spin Chern number directly in the model we consider because it conserves the $z$-component of the spin.\cite{prb0890235104}

\begin{figure*}[th]
\centering
\includegraphics[width=17.0cm]{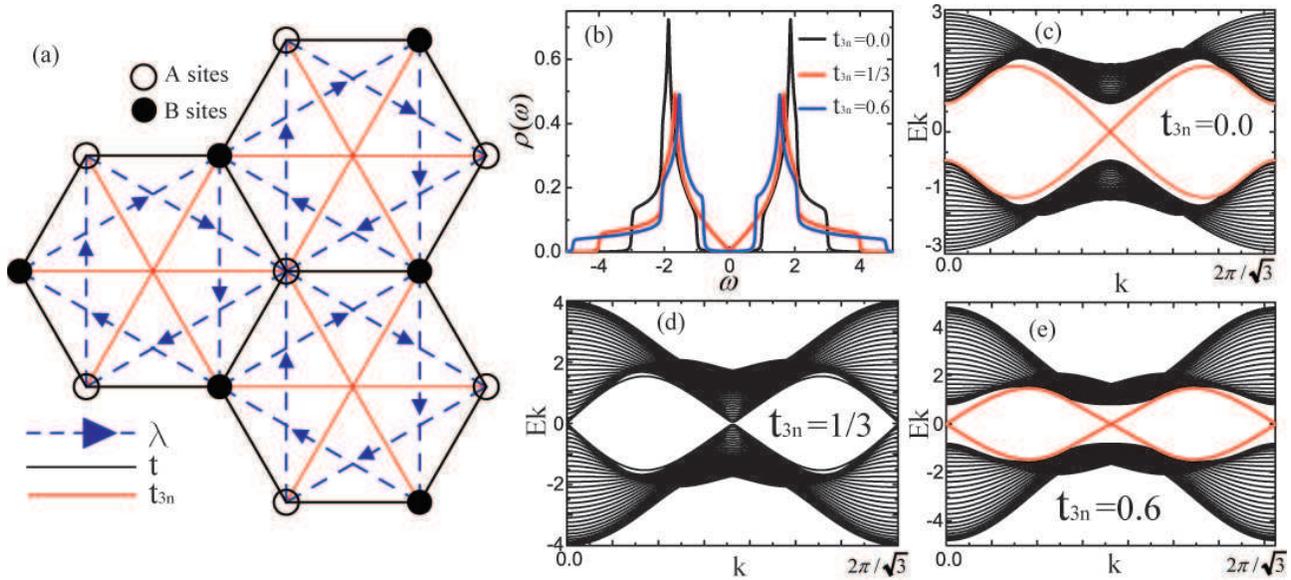}\hspace{0.5cm}
\caption{\label{fig:dosekwithU0}(Color online) (a) The lattice structure of honeycomb lattice with various hopping parameters described in Eq. (\ref{eq:hamilton}). (b) The bulk density of states for various values of third-neighbor hopping, $t_{3n}$, for $\lambda=0.4, U=0.0$. (c)-(e) The noninteracting energy bands of Eq. (\ref{eq:hamilton}) with the armchair strip geometry for (c) $t_{3n}=0.0$, (d) $t_{3n}=1/3$, and (e) $t_{3n}=0.6$. The width is  $N_s=120$ and spin-orbit coupling is $\lambda=0.4$. A single Dirac cone is present in (c), indicating a topological state, while two are present in (e), a trivial state.  The bulk band gap is closed in (d), as in (b) for $t_{3n}=1/3$.
}
\end{figure*}

In this work, we investigate topological phase transitions in an interacting honeycomb lattice model (a generalized Kane-Mele-Hubbard model--see below) with a third-neighbor hopping, $t_{3n}$, using CDMFT with the CTQMC solver.  At weak interactions, we find a gapped topological band insulator with spin Chern number $|C|=1$ when $t_{3n}=0$ that persists until $t_{3n}=t^c_{3n}=t/3$, where a gapless metallic state appears. The bulk gap is reopened and spin Chern number $|C|=2$ when $t_{3n}$ is increased, indicating the system becomes a topologically trivial state (TTI)\cite{prb0890235104}.  Combined with the recent proposal to detect a (spin) Chern number variation in a two-level  system via a superconducting qubit\cite{prl113050402},  it appears that topological phase transitions may be directly observed experimentally. In addition to a change in the topological invariants, topological phase transitions can also be signaled by a gap closing.

In contrast to the zero temperature case\cite{ prb0870121113r}, the gapless Dirac cone structures at the topological phase boundary give rise to a finite-ranged (in terms of third-neighbor hopping) metallic state at finite temperatures and finite interactions. This intermediate phase exhibits a spin Hall effect. Under strong interactions, an xy-easy plane anti-ferromagnetic insulating state is observed for all values of $t_{3n}$ when the interaction $U$ is beyond a critical value. Thus, there is no topological phase transition in the strong coupling limit. One of our main results is the finite temperature phase diagram Fig.\ref{fig:phase_diagram}, given in terms of the interaction strength and the third neighbor hopping. These interesting phases could be experimentally probed by transport, angle-resolved photoemission spectroscopy (ARPES) \cite{rmp0750473}, neutron scattering, nuclear magnetic resonance (NMR) \cite{prb0650144447}, and other experiments. For interaction values below the critical strength required to induce a magnetic transition, our results are in good quantitative agreement with recent fermion-sign free quantum Monte Carlo calculations on the same model\cite{prb0890235104}. (The QMC study did not explore the strong coupling regime.)

Our paper is organized as follows. In Sec. \ref{sec:model}, we introduce the interacting honeycomb lattice model we study, and the cellular dynamical mean field theory. In Sec. \ref{sec:results}, we present the main results of our CDMFT study, including the spectral function, explicit computations of the edge-state spectrum, and the dependence of various excitations gaps on the parameters of the Hamiltonian.  Finally, in Sec. \ref{sec:phase_diagram} we present a finite-temperature phase diagram of our model, which includes an anti-ferromagnetic phase, a topological insulating phase, and a trivial insulating phase.  We summarize our results in Sec. \ref{sec:summary}.


\section{Model and Method}
\label{sec:model}

We consider the standard Kane-Mele-Hubbard model at half-filling (one electron per site) on the honeycomb lattice:
\begin{eqnarray}\label{eq:hamilton}
&&\nonumber H=-t\sum_{\langle{ij}\rangle\sigma}c_{i\sigma}^{+}c_{j\sigma} +i\lambda\sum_{\langle\langle{ij}\rangle\rangle}c_{i\sigma}^{+}v_{ij}(\sigma)c_{j\sigma} \\
&&\nonumber -t_{3n}\sum_{\langle\langle\langle{ij}\rangle\rangle\rangle\sigma}c_{i\sigma}^{+}c_{j\sigma}+h. c.\\
&&+U\sum_{i}n_{i\uparrow}n_{i\downarrow}+\mu\sum_{i\sigma}n_{i\sigma},
\end{eqnarray}
where $t$ is the nearest-neighbor (NN) hopping energy, $\lambda$ is the spin-orbit coupling strength, $v_{ij}(\sigma)$ takes opposite signs for different spin projections and depends on the second-neighbor bond $\langle\langle{ij}\rangle\rangle$\cite{prl0950146802,prl0950226801}, $t_{3n}$ is the next-next-nearest-neighbor (NNNN) hopping energy, $U$ is the on-site repulsive interaction, $\mu$ is the chemical potential which keeps the system at half filling, $c_{i\sigma}^{+}$ and $c_{i\sigma}$ denote the creation and annihilation operators respectively, $n_{i\sigma}=c_{i\sigma}^{+}c_{i\sigma}$ corresponds to the density operator, and $\sigma$ runs over spin up ($\uparrow$) and spin down ($\downarrow$). Here, we set $t=1.0$, which is also used as the energy unit in our paper. The spin-orbit coupling strength $\lambda$ is taken to be $\lambda=0.4$.

The lattice structure is shown in Fig. \ref{fig:dosekwithU0} (a). The honeycomb lattice can be divided to two sublattices, designated by A sites and B sites. In Fig. \ref{fig:dosekwithU0} (a), the A sites are denoted by white circles, and the black circles shows the B sites. The NN hopping $t$ is shown by the black solid lines, while the blue dash lines describe the spin-orbital coupling strength. The NNNN hopping is demonstrated by the red solid lines. The bulk density of states for various $t_{3n}$ when $\lambda=0.4$ are shown in Fig. \ref{fig:dosekwithU0} (b). A visible bulk gap opened by the spin-orbit coupling is found when $t_{3n}$ is absent. This bulk gap is closed for $t_{3n}=1/3$, independent of the value of $\lambda$. The gapless behavior means that the system becomes a metal.  In contrast to graphene, in the  generalized Kane-Mele model, the Driac cones are located at three time-reversal invariant momenta $M_{1,2}=(\pm \frac{\pi}{\sqrt{3}}, \frac{\pi}{3})$ and $M_3=(0,\frac{2\pi}{3})$\cite{ prb0870121113r,prb0890235104}. The bulk gap is reopened when $t_{3n}>1/3$, such as for $t_{3n}=0.6$. This gapped-gapless-gapped behavior indicates that a topological phase transition may be found when $t_{3n}$ is  tuned, and the topological phase boundary is $t^c_{3n}=1/3$.  A direct evaluation of the topological invariant, and band structure computations in a strip geometry confirm this is indeed the case.

Fig. \ref{fig:dosekwithU0} (c)-(d) shows the energy bands in a strip geometry for different $t_{3n}$ when $\lambda=0.4$, which is obtained with an armchair boundary condition. The presence of an odd number of helical edge states (with time-reversed spins) is characteristic of the nontrivial TBI\cite{prl0960106401}. Clear edge states crossing the bulk gap with one Dirac point are found for $\lambda=0.4, t_{3n}=0.0$ in Fig. \ref{fig:dosekwithU0} (c), implying that the system is a topological band insulator, and the spin Chern number $C_{\sigma}=\pm 1$. Upon increasing $t_{3n}$ to $t^c_{3n}=1/3$,  both the edge and bulk states become gapless (see Fig. \ref{fig:dosekwithU0} (d)), indicating that the system is a metal (M). When $t_{3n}=0.6$, the bulk gap reopens and edge states with two Dirac points appear (see Fig. \ref{fig:dosekwithU0} (e)), showing that the state is a topological trivial insulator (TTI). In the trivial case, the spin Chern number $C_{\sigma}=\pm 2$. As long as $S_z$ is conserved, the spin Chern number is a good quantity to describe the topological properties.

In order to address the Hubbard interaction term in the model given in Eq.\eqref{eq:hamilton}, we use CDMFT with the CTQMC solver to investigate the topological and magnetic phase transitions on the honeycomb lattice with NNNN (third neighbor, $t_{3n}$) hopping. In CDMFT,
we map the original lattice model onto an effective cluster model coupled to an effective medium via a standard dynamical mean-field theory (DMFT) procedure. The single-particle Green's function of the cluster, $\hat{g}$, in the effective medium is obtained from
\begin{eqnarray}
\hat{g}^{-1}(i\omega)=\Big(\sum_{\vec{k}}\frac{1}{i\omega+\mu-\hat{t}(\vec{k})-\hat{\Sigma}(i\omega)}\Big)^{-1}+\hat{\Sigma}(i\omega),
\end{eqnarray}
where $\hat{t}(\vec{k})$ is the hopping matrix of the original model Hamiltonian, $\vec{k}$ is the wave vector within the reduced Brillouin zone based on the cluster size and geometry, $\hat{\Sigma}(i\omega)$ is the self-energy, and $\omega$ is the Matsubara frequency. The matrix $\hat{g}$ can be used as an input to an impurity solver, such as CTQMC, to obtain the Green's function $\hat{G}(i\omega)$ of the physical problem of interest. The new self-energy $\hat{\Sigma}(i\omega)$ is obtained via the Dyson equation $\hat{\Sigma}(i\omega)=\hat{g}^{-1}(i\omega)-\hat{G}^{-1}(i\omega)$ to close the self-consistent iterative loop. This loop is repeated until the self-energy $\hat{\Sigma}(i\omega)$ converges to the desired accuracy. In this paper, we use $N_c=8$ ($N_c$ is the cluster size) in the CDMFT calculation.  The interacting edge spectra are obtained by the momentum-dependent spectral function for the armchair strip geometry with the width of  $N_s=80$.

\begin{figure}[t!]
\centering
\includegraphics[width=9.0cm]{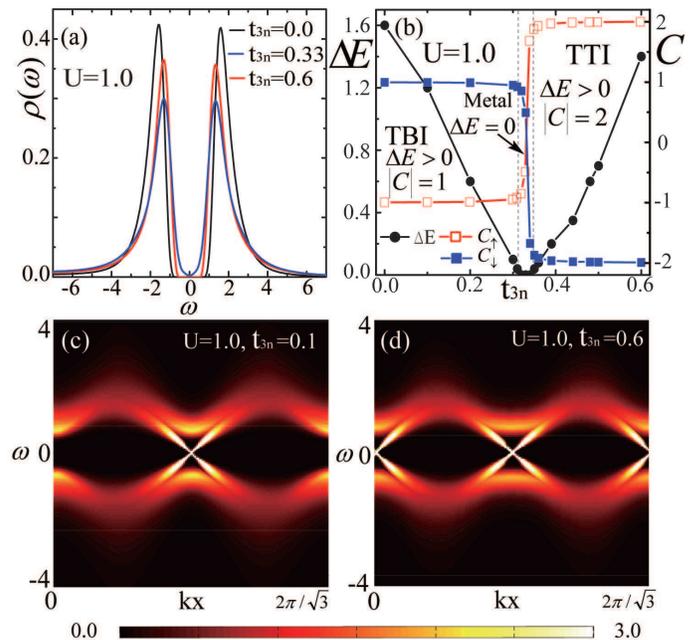}\hspace{0.5cm}
\caption{\label{fig:dosscafw}(Color online) (a) The bulk density of states for different $t_{3n}$ when $\lambda=0.4, T=0.05, U=1.0$. (b)The evolution of the single-particle gap ${\Delta}E$ (left vertical axis) and spin Chern number $C_{\sigma}$ (right vertical axis) as a function of $t_{3n}$ when $\lambda=0.4, T=0.05, U=1.0$. (c)-(d) The momentum-dependent spectral function, $A(k,\omega)$, in a strip geometry for (c) $t_{3n}=0.0$ and (d) $t_{3n}=0.6$ when $\lambda=0.4, T=0.05, U=1.0$.  By a comparison with Fig.\ref{fig:dosekwithU0}, it is clear that (c) is a TBI and (d) is a TTI. 
}
\end{figure}

The spin Chern number can be obtained by the Green's function at zero frequency and projection operator formalism, the details of which can be found in Ref.~[\onlinecite{prb0890235104}]. With the Matsubara frequency Green's functions, we can perform an analytical continuation to obtain real frequency Green's functions using the so-called Maximum Entropy Method (MEM) \cite{Gubernatis}. The density of states $\rho(\omega)$ as well as the single-particle gap $\Delta E$ can be described in terms of the spectral functions $A(\omega)=-\frac{1}{\pi}\textrm{Im}G(\omega+i\delta)$, where $\delta$ is a positive infinitesimal.

 At strong interactions, an easy-plane anti-ferromagnetic state develops, and is observed in CDMFT by introducing a symmetry-breaking perturbation in Eq. (\ref{eq:hamilton}). The Neel temperature $T_N$ decreases as the cluster size $N_c$ increases, eventually tending towards zero\cite{prl0950237001}. Therefore, in the two-dimensional systems we consider, the Mermin-Wigner theorem is recovered as $N_c{\rightarrow}{\infty}$.  Thus, the anti-ferromagnetic state of Eq. (\ref{eq:hamilton}) disappears when $N_c{\rightarrow}{\infty}$ at finite temperatures.



\section{Results}
\label{sec:results}

\subsection{Parameter-driven topological phase transition}

We first focus on the topological phase transition driven by the NNNN hopping, $t_{3n}$, at weak interactions. Fig. \ref{fig:dosscafw} (a) shows the evolution of the density of states (DOS) for different values of $t_{3n}$ when $\lambda=0.4, U=1.0, T=0.05$. Similar to the non-interacting situation, the bulk gap induced by the spin-orbit coupling is present when $t_{3n}=0.0$. This gap is closed when $t_{3n}=0.33$, indicating a metallic behavior. Moreover, the gap is re-opened when $t_{3n}$ is increased. This gapped-gapless-gapped behavior indicates that a topological phase transition can be found when $t_{3n}$ is increased from zero, and this is confirmed by investigating the spin Chern number and the edge modes in a strip geometry. In Fig. \ref{fig:dosscafw} (b), we show the development of the single-particle gap ${\Delta}E$ and spin Chern number $C_\sigma$ as a function of $t_{3n}$ for $\lambda=0.4, U=1.0, T=0.05$, where $\sigma$ denotes the spin. The single-particle gap ${\Delta}E$ decreases when the $t_{3n}$ is increased towards 1/3, and the $|C_\sigma|=1$ character is maintained. When $0.32<t_{3n}<0.35$, the  ${\Delta}E$ is decreased to zero, indicating a metallic state.
When $t_{3n}>0.36$, the bulk gap is reopened (${\Delta}E{\neq}0$) and $|C_\sigma|=2.0$, indicating that the system becomes a $Z_2$ TTI.
Note that, in the $\Delta E=0$ regime, the spin Chern numbers Fig. \ref{fig:dosscafw} (b) are not described, since the topological invariants are not well defined in the metallic state.

The finite parameter extent of the metallic state shown in Fig. \ref{fig:dosscafw} (b) is a finite temperature effect. At zero temperature, the sign-free QMC study shows a line-like  topological phase boundary\cite{prb0870121113r} when $t_{3n}=t^c_{3n}$. This is because at half-filling the Fermi surface is point-like at $t^c_{3n}$, with the Dirac points at $M_{1,2,3}$. However, at finite temperatures, thermal fluctuations smear the distribution of electronic states away the Dirac points. Thus the metallic state can be extended to a finite range of $t_{3n}$. This behavior has also been observed in graphene\cite{prb0850205102}. The thermal-fluctuation-induced metallic state exhibits a spin Hall effect. We find that the time-reversal symmetry is still present in this regime, and the presence of spin-orbit coupling in Eq. (\ref{eq:hamilton}) will bring a spin accumulation on the edges \cite{prl08301834, prl0990126601}.

Even with finite interactions, the helical edges states remain characteristic of the nontrivial topology.  In order to check whether the system is truly a topological insulating state for $t_{3n}<0.31$ and $t_{3n}>0.36$, we obtain the momentum-dependent spectral function for a strip geometry with an armchair boundary condition ($N_s=80$) in Fig. \ref{fig:dosscafw} (c) and (d).  Clear edge states with one Dirac point are found in Fig. \ref{fig:dosscafw} (c), indicating $|C_\sigma|=1.0$ when $t_{3n}=0.1$ for $\lambda=0.4, U=1.0, T=0.05$. In Fig. \ref{fig:dosscafw} (d), we  find edge states with two Dirac points, meaning that $|C_\sigma|=2.0$. These results are consistent with the evolution of $|C_\sigma|$ as a function of $t_{3n}$, as shown in Fig. \ref{fig:dosscafw} (b).

\begin{figure}[t]
\centering
\includegraphics[width=9.0cm]{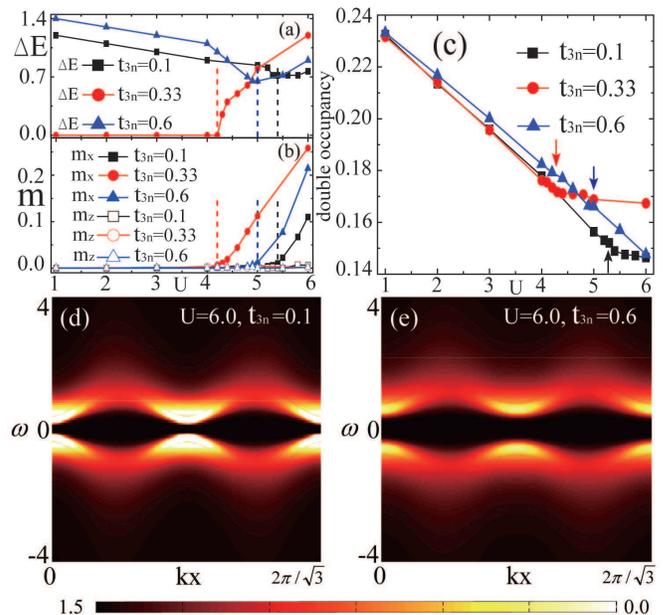}\hspace{0.5cm}
\caption{\label{fig:makwdocc}(Color online) The evolution of (a) bulk single particle gap ${\Delta}E$, (b) staggered magnetic moment $m_z$ and transverse magnetism $m_x$ as a function of interaction $U$ when $\lambda=0.4, T=0.05$ for different $t_{3n}$. The dashed lines show the critical interaction $U_c$, which depends on $t_{3n}$. (c) The evolution of double occupancy as a function of $U$, the arrows show the critical points for different $t_{3n}$ when $\lambda=0.4, T=0.05$. (d) - (e): The momentum-dependent spectral function, $A(k,\omega)$, obtained from the armchair boundary condition for (d) $t_{3n}=0.1$, (e) $t_{3n}=0.6$ when $\lambda=0.4, U=6.0, T=0.05$.  Note that $U=6.0>U_c$ and the spectrum is fully gapped. }
\end{figure}

\subsection{Interaction-driven topological phase transition}

Next, we turn to study the topological phase transitions driven by the Hubbard interaction $U$. In order to study a possible magnetic phase transition in the system, we measure the staggered diagonal magnetic moment $m_z$ and the staggered transverse magnetic moment $m_x$. The diagonal magnetic moment is defined as $m_z \equiv \frac{1}{N}\sum_{i}^{N}sgn(i)(n_{i\uparrow}-n_{i\downarrow})$, with $N$ denotes the number of sites in the lattice, $i$ means the site index shown in Fig.~\ref{fig:dosekwithU0} (a), $sgn(i)=+1$ for $i$ corresponding to A sites, and $sgn(i)=-1$ for $i$ corresponding to B sites. The transverse magnetism $m_x$ is defined as $m_x \equiv \frac{1}{N}\sum_{i=1}^{N} sgn(i)\langle S_i^x \rangle=\frac{1}{N}\sum_{i=1}^{N}\frac{1}{2} sgn(i) \langle c^{+}_{i\uparrow}c_{i\downarrow}+h.c.\rangle$.

In Fig.~\ref{fig:makwdocc} (a), we show the evolution of the bulk single-particle gap ${\Delta}E$ as a function of interaction $U$ for various $t_{3n}$. We find that for $\lambda=0.4, t_{3n}=0.1, 0.6$,  ${\Delta}E$ decreases when the interaction $U$ is increased. This means that the bulk gap induced by the spin-orbital coupling is suppressed by the interaction.  Note, however, that the single-particle gap {\em does not close} across the critical value of interactions where the system become magnetic. This is consistent with QMC results obtained on the Kane-Mele-Hubbard model\cite{prl0106100403,prb085115132}.  Different from the cases of $t_{3n}=0.1, 0.6$, a gapless behavior is found when $U<U_c=4.3$ for $t_{3n}=0.33$, implying that the system is a metal over a range of interaction strengths. When $U>4.3$ for $t_{3n}=0.33$, a bulk gap is opened by the interaction and the system becomes an insulator.

The development of $m_z$ and $m_x$ at various values of $U$ is shown in Fig.~\ref{fig:makwdocc} (b), in which the dashed lines show the critical points $U_c$ for different $t_{3n}$.  When $U>U_c$, the $m_x$ is increased to a finite value while the $m_z$ remains zero. This indicates a phase transition from a paramagnetic state to an anti-ferromagnetic insulating state, which was also found in Kane-Mele-Hubbard model studies \cite{prl0106100403,prb085115132,prl0107010401, prb0850205102}.
The finite $m_x$ means that this magnetic order is formed in the easy-plane. Fig. ~\ref{fig:makwdocc} (c) shows the evolution of double occupancy $d_{occ}$, which is defined as $d_{occ}=\frac{\partial{F}}{\partial{U}}=\frac{1}{N}\sum_{i=1}^{N}\langle{n_{i\uparrow}n_{i\downarrow}}\rangle$, where $F$ denotes the free energy. The $d_{occ}$ can be used to check the phase transition order because it is directly connected to the free energy. In Fig.~\ref{fig:makwdocc} (c), we find that $d_{occ}$ decreases when the interaction $U$ is increased, indicating that the itinerancy of the particles is suppressed by the interaction. The arrows in Fig ~\ref{fig:makwdocc} (c) shows the critical points for different $t_{3n}$. At the same corresponding $U$s, the magnetic moments develop in Fig. ~\ref{fig:makwdocc} (b). The high $d_{occ}$ in weak interaction indicates that when $U<U_c$ the observed bulk gap is not induced by the interaction. The smooth decreasing of $d_{occ}$ means this magnetic phase transition is a second order  phase transition.

 The presence of an spontaneous easy-plane anti-ferromagnetic order above a critical interaction strength mixes the spin components. Thus, $S_z$ is no longer conserved, spin Chern numbers are not quantized, and there are gapless Goldstone modes in the spin channel. To further examine the topological properties of the anti-ferromagnetic insulating states, we study the edge states of Eq. (\ref{eq:hamilton}) at strong coupling. Fig.~\ref{fig:makwdocc} (d) and (e) show the momentum-dependent spectral function, $A(k,\omega)$, obtained for the strip geometry with an armchair boundary condition.  In contrast to the weak interaction situation, no edge state is found for $t_{3n}=0.1$ and $0.6$ when $U=6.0, T=0.05, \lambda=0.4$. This means that across all the values of $t_{3n}$, the topological state is destroyed by the strong interaction, and both of the TBI and TTI turn to the trivial easy-plane anti-ferromagnetic insulating state. As a consequence, there is no topological phase transition in the strong coupling limit.


\section{Phase Diagram}
\label{sec:phase_diagram}

\begin{figure}[t]
\centering
\includegraphics[width=9.0cm]{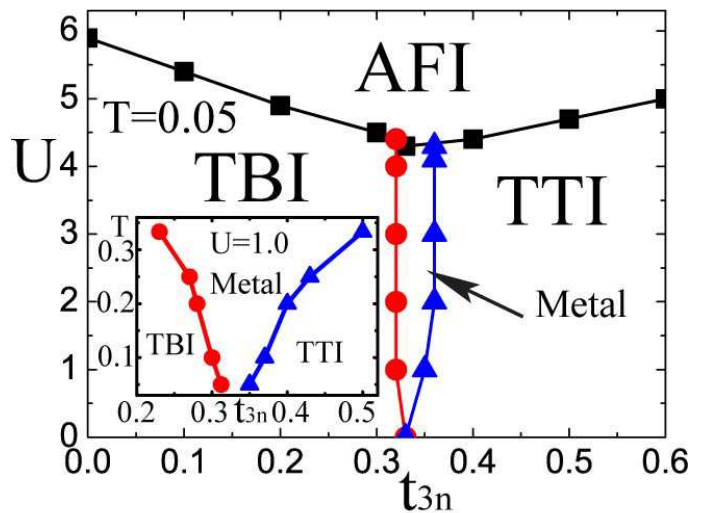}\hspace{0.5cm}
\caption{\label{fig:phasediagram}(Color online) The $U-t_{3n}$ phase diagram for $T=0.05$.  $\lambda=0.4$ is considered. TBI: topological band insulator; TTI: topological trivial insulator; AFI: anti-ferromagnetic insulator; M: metal. The inset shows the temperature dependence of the metallic state at $U=1.0, \lambda=0.4$. }
\label{fig:phase_diagram}
\end{figure}

Using CDMFT, we investigate the finite temperature effects in the Kane-Mele-Hubbard model with third-neighbor hopping $t_{3n}$. The phase diagram as a function of $t_{3n}$ and $U$ is summarized in Fig.~\ref{fig:phasediagram}. When $U<U_c$, a TBI-M-TTI phase transition occurs when $t_{3n}$ is increased. The TBI phase can be identified by ${\Delta}E\neq0, |C_{\sigma}|=1$, and the TTI state can be found by ${\Delta}E\neq0, |C_{\sigma}|=2$, as well as by the different edge states in the TBI and TTI. A metallic state can be found when $t_{3n}^{C_{TBI-M}}<t_{3n}<t_{3n}^{C_{M-TTI}}$ with ${\Delta}E=0$.   Due to the presence of spin-orbit coupling and time-reversal symmetry, at sufficiently low temperature, the metallic state exhibits a spin Hall effect. The inset of Fig.~\ref{fig:phasediagram} shows that this metallic state is enlarged when the temperature is increased, but we expect that beyond certain temperatures the spin Hall state effect will vanish.

An easy-plane anti-ferromagnetic insulating state can be found when the interaction is increased, such as $U>U_c=4.3, t_{3n}=0.33$. A clear gap can be found in the easy-plane anti-ferromagnetic insulating state, in which $m_z$ remains at zero, and $m_x$ is increased to a finite value. The topological property of the magnetic state is further reexamined by studying the edge states using CDMFT and the maximum entropy method. We do not find any coexisting region of a topological state and an anti-ferromagnetic insulating state.

\section{Summary}
\label{sec:summary}

In summary, we have studied the Kane-Mele-Hubbard model with an additional third neighbor hopping term at finite temperature using cellular dynamical mean-field theory with a continuous-time quantum Monte Carlo impurity solver. The third-neighbor hopping on the honeycomb lattice with spin-orbit coupling can induce a topological phase transition to a trivial state for small interaction values.  A metallic state with a vanishing single particle gap ${\Delta}E=0$ is found in a small region of third-neighbor hopping for interaction values below a certain critical strength. When the interaction is stronger than the critical interaction, an easy-plane anti-ferromagnetic
insulating state with transverse magnetic order is formed.  The same magnetic state is found ``above" the topological trivial state and the topological non-trivial state.  In addition, we have presented the spectral function of the system for various Hamiltonian parameters through a parameter space representative of the full phase diagram.
Our study provides an important step for understanding finite temperature effects on the topological phase transition to magnetic order in spin-orbit coupled systems. In our study, we did not find any coexisting state of the magnetic order and the $Z_2$ topological order. In recently years, a novel correlated material, Na$_2$IrO$_3$, has emerged as a good candidate to investigate the phase transition induced by the interaction and spin-orbit coupling. The relative strength of the  NNNN hopping can in principle be adjusted by physical and chemical pressure.

We would like to thank Fadi Sun, Yuan-Yen Tai, and Yuan-Yuan Zhao for valuable
discussions. This work was supported by the Texas Center for Superconductivity at the University of Houston and by the Robert A. Welch Foundation under Grant No. E-1146. HHH and GAF acknowledge financial support through ARO Grant No. W911NF-09-10527, NSF Grant No. DMR-0955778, and DARPA Grant No. D13AP00052. HHH also thanks Emanuel Gull and Hua Chen for insightful the discussions.

\end{document}